\documentclass[aps,floats,floatfix,showpacs,preprintnumbers,amssymb,prd,twocolumn,superscriptaddress,nofootinbib,nolongbibliography,reprint]{revtex4-1}

\usepackage{amssymb,amsmath,verbatim,mathtools,needspace,enumitem,etoolbox,graphicx,physics,microtype,afterpage,xspace,tabularx,lmodern,multirow,boldline, makecell, booktabs}

\usepackage{appendix}
\usepackage{tensor}
\usepackage{array}
\usepackage{siunitx}
\usepackage[normalem]{ulem}
\usepackage[dvipsnames, usenames]{xcolor}
\usepackage{xr-hyper}
\usepackage[font=small,labelfont=bf]{caption}
\definecolor{linkcolor}{rgb}{0.0,0.3,0.5}
\usepackage[unicode, colorlinks=true, linkcolor=linkcolor, citecolor=linkcolor, filecolor=linkcolor, urlcolor=linkcolor, linktocpage, breaklinks]{hyperref}
\usepackage[all]{hypcap}
\usepackage[T1]{fontenc}
\usepackage[utf8]{inputenc}
\usepackage[usenames,dvipsnames]{xcolor}
\definecolor{cornellGreen}{HTML}{6EB43F}
\definecolor{cornellRed}{HTML}{B31B1B}
\hypersetup{colorlinks=true,citecolor=cornellRed,linkcolor=cornellRed,urlcolor=cornellRed}
\usepackage{multirow}

\definecolor{romared}{RGB}{142,0,28}

\newcommand{\be}{\begin{equation}}
\newcommand{\ee}{\end{equation}}

\def\be{\begin{equation}}
\def\ee{\end{equation}}
\newcommand{\beq}{\begin{eqnarray}}
\newcommand{\eeq}{\end{eqnarray}}

\usepackage{makecell}
\usepackage{soul}

\newcolumntype{Y}{>{\centering\arraybackslash}X}

\usepackage{graphicx}
\usepackage{dcolumn}
\usepackage{bm}
\usepackage{hyperref}
\usepackage{color}

\usepackage{times} 

\usepackage{graphicx} 
\usepackage{booktabs} 
\usepackage[font=small,labelfont=bf]{caption} 
\usepackage{amsfonts, amsmath, amsthm, amssymb} 

\usepackage{wrapfig} 
 \definecolor{mypurple}{RGB}{130, 0, 130} 

\begin{document}
\title{Root‑$T\bar{T}$ Flows Unify 4D Duality-Invariant Electrodynamics and 2D Integrable Sigma Models}
\author{H. Babaei-Aghbolagh}
\email{hosseinbabaei@nbu.edu.cn}
\affiliation{Institute of Fundamental Physics and Quantum Technology, \& School of Physical Science and Technology, Ningbo University, Ningbo, Zhejiang 315211, China}
\author{Bin Chen}
\email{chenbin1@nbu.edu.cn}
\affiliation{Institute of Fundamental Physics and Quantum Technology, \& School of Physical Science and Technology, Ningbo University, Ningbo, Zhejiang 315211, China}
\affiliation{School of Physics, Peking University, \& Center for High Energy Physics,  No.5 Yiheyuan Rd, Beijing
100871, P.R. China}
\author{Song He}
\email{Corresponding author: hesong@nbu.edu.cn}
\affiliation{Institute of Fundamental Physics and Quantum Technology, \& School of Physical Science and Technology, Ningbo University, Ningbo, Zhejiang 315211, China}
\affiliation{Max Planck Institute for Gravitational Physics (Albert Einstein Institute), Am M\"uhlenberg 1, 14476 Golm, Germany}

\date{\today}
\begin{abstract}
We present a unified framework that connects four-dimensional duality-invariant nonlinear electrodynamics and two-dimensional integrable sigma models via the Courant–Hilbert and new auxiliary field formulations, both governed by a common generating function and a generating potential, respectively. Introducing two commuting deformation parameters, $\lambda$ (irrelevant) and $\gamma$ (marginal), we identify a universal class of $\gamma$-flows, including the root-$T\bar{T}$ deformation and its rescaled variants. Our approach generalizes conventional single-coupling structures via novel field transformations that extend to a two-parameter space ($\lambda$,$\gamma$) while preserving the root-$T\bar{T}$ flow condition for all $\gamma$-coupled theories. We construct several integrable models, including generalized Born–Infeld, logarithmic, q-deformed, and a new closed-form theory applicable to both electrodynamics and integrable systems. This unified framework, based on the unique form of the root-$T\bar{T}$ flow, systematically spans duality-invariant nonlinear electrodynamics in 4D and their exact 2D integrable counterparts.
\end{abstract}
\maketitle

\section*{Introduction}
Understanding the connection between \emph{self‑duality} and \emph{integrability} in field theories has long been a crucial challenge in theoretical physics. The self-duality condition in four-dimensional (4D) nonlinear electrodynamics (NLED) constrains the form of the electromagnetic dynamics, dictating a highly specific class of interactions that are both causal and UV-tame, and arising naturally in D-brane and string effective actions~\cite{Born:1934gh, Fradkin:1985qd, Garousi:2017fbe}. In parallel, integrability in two-dimensional (2D) models plays a significant role in offering exact solutions, nonperturbative control through Lax pairs, and a wealth of conserved charges~\cite{Faddeev:1987ph}. 



Both integrability and duality-invariant conditions manifest as partial differential equations (PDEs)~\cite{Bialynicki-Birula:1984daz, Gaillard:1997rt, Gibbons:1995cv, Gibbons:1995ap, Ferko:2023pr, Murcia:2025psi}, which constrain the admissible forms of nonlinear dynamics in two and four dimensions. Notably, several case studies have shown that $T\bar{T}$-type deformations can preserve both integrability in two-dimensional sigma models and electromagnetic self-duality in four-dimensional theories~\cite{Smirnov:2016lqw, Cavaglia:2016oda, Conti:2018jho, BabaeiAghbolagh:2025rtt, Chen:2025ndc}. In two dimensions, the corresponding flow equations for scalar field theories with integrability conditions under $T\bar{T}$ deformations were derived in~\cite{Smirnov:2016lqw, Cavaglia:2016oda}. Analogously, for four-dimensional duality-invariant electrodynamics~\cite{Bialynicki-Birula:1984daz}, flow equations were first proposed in~\cite{Conti:2018jho} and subsequently extended in~\cite{Babaei-Aghbolagh:2020kjg, Ferko:2019oyv, Babaei-Aghbolagh:2024uqp}. However, these results were obtained on a case-by-case basis, lacking a unified formalism that systematically captures the parallel structure between integrability in 2D and self-duality in 4D. A general framework that treats both sectors within a common deformation paradigm has remained elusive until now.

In this Letter, we unify two types of deformations, irrelevant ($\lambda$) and marginal ($\gamma$), into a single, dimension-independent framework. This is achieved by combining the Courant–Hilbert (CH) characteristic method~\cite{Courant:Hilbert} with a new auxiliary-field formulation~\cite{Lechner:2022qhb,Russo:2025fuc}. The resulting construction applies uniformly to 2D integrable sigma models and to 4D duality-invariant electrodynamics, and it identifies the precise rescalings of the root-$T\bar{T}$ flow that preserve both duality invariance and integrability.

Our approach also extends the marginal $\gamma$ flows~\cite{Babaei-Aghbolagh:2022uij}, originally appearing in ModMax theory~\cite{Bandos:2020jsw}, to a broad class of $\lambda$-coupled single-deformation theories, including Born–Infeld, logarithmic and q-deformed  models \cite{Russo:2024csd,Russo:2024dual,Russo:2024xnh}. By deforming the generating function within both the CH formalism and the auxiliary-field perspective, we derive new closed-form solutions that were previously unknown. These results provide a systematic framework for embedding marginal flows into known irrelevant deformations and represent a key advancement introduced here for the first time.

\section*{Theory and Method}
Duality‐invariant deformations of the Maxwell theory in four dimensions are encoded in a Lagrangian $\mathcal{L}(S, P)$, where
\begin{equation}
S=-\tfrac14F_{\mu\nu}F^{\mu\nu}, 
\quad
P=-\tfrac14F_{\mu\nu}\widetilde F^{\mu\nu}.
\end{equation}
Consistency under electric–magnetic rotations imposes the self‐duality condition~\cite{Bialynicki-Birula:1984daz, Gaillard:1997rt, Gibbons:1995cv, Gibbons:1995ap}
\begin{equation}
\bigl(\mathcal{L}_S\bigr)^2
-2\,\frac{S}{P}\,\mathcal{L}_S\,\mathcal{L}_P
-\bigl(\mathcal{L}_P\bigr)^2
=1,
\end{equation}
with $\mathcal{L}_{S,P}\equiv\partial\mathcal{L}/\partial(S,P)$.  Defining
\begin{equation}
U=\frac{1}{2}\bigl(\sqrt{S^2+P^2}-S\bigr), 
\quad
V=\frac{1}{2}\bigl(\sqrt{S^2+P^2}+S\bigr),
\end{equation}
one recasts the duality‐invariant condition in the simple form~\cite{Gaillard:1997rt}:
\begin{equation}\label{PDE4UV}
\mathcal{L}_U\,\mathcal{L}_V=-1.
\end{equation}
Two-dimensional integrable sigma models with group-valued fields 
$g(\sigma,\tau)$ are characterized by the flatness of a Lax connection. In light-cone coordinates, define the
currents $j = g^{-1}dg, \quad j_{\pm} = j_0 \pm j_1$ and the scalar invariants
\begin{equation}
P_1 = -\text{tr}[j_+ j_-], \quad
P_2 = \tfrac12\left( \text{tr}[j_+ j_+]\,\text{tr}[j_- j_-] + (\text{tr}[j_+ j_-])^2 \right).
\end{equation}
The Lagrangian $\hat{\mathcal{L}}(P_1, P_2)$ must satisfy the PDE~\cite{Borsato:2022tmu}:
\begin{equation}
8(P_1^2 - P_2)\,\hat{\mathcal{L}}_{P_2}^2
+ 8P_1\,\hat{\mathcal{L}}_{P_2}\,\hat{\mathcal{L}}_{P_1}
+ 4\,\hat{\mathcal{L}}_{P_1}^2 = 1.
\end{equation}
Introducing
\begin{equation}
q_{1} = \tfrac14\big( \sqrt{2P_2 - P_1^2} - P_1 \big), \, q_{2} = \tfrac14\big( \sqrt{2P_2 - P_1^2}+ P_1 \big),
\end{equation}
this reduces to a simple condition:
\begin{equation}\label{PDE222}
\hat{\mathcal{L}}_{q_1}\, \hat{\mathcal{L}}_{q_2} = -1.
\end{equation}

Equations~\eqref{PDE4UV} and~\eqref{PDE222} are the same first–order nonlinear PDE~\footnote{Since PDEs~\eqref{PDE4UV} and~\eqref{PDE222} share the same form, their solutions can be derived similarly. We adopt the notation of two-dimensional scalar fields with $\hat{\mathcal{L}}(q_1,q_2)$, which can be translated into duality-invariant electromagnetic theories using the $(q_1,q_2) \to (U, V)$ dictionary. Using this dictionary, you can get: $\hat{\mathcal{L}}(q_1,q_2) \to \mathcal{L}(U, V)$.}, so four–dimensional self-duality and two-dimensional classical integrability are controlled by one universal characteristic equation. Hence, both sectors may be solved uniformly by a single generating CH-function $\ell(\tau)$ introduced in the next section, which simultaneously encodes marginal ($\gamma$) and irrelevant ($\lambda$) stress–tensor flows in either dimension.

\subsection*{Courant–Hilbert framework}

At the heart of the Courant–Hilbert framework~\cite{Courant:Hilbert} is a single generating function, $\ell(\tau)$, that solves both the four-dimensional duality-invariant and two-dimensional integrability conditions. 
For each pair in the positive quadrant, we need to define $\hat{\mathcal{L}}(q_1, q_2)$. Subject to $\hat{\mathcal{L}}(0,q_2)=\ell(q_2)$, is given by the characteristic ansatz:
\begin{equation}\label{Soul}
\hat{\mathcal{L}}\,=\,\ell(\tau)\;-\;\frac{2\,q_1}{\ell^{\prime}(\tau)},
\quad
\tau \;=\;q_2 + \frac{q_1}{\ell^{\prime}(\tau)^2},
\end{equation}
where $\ell(\tau)$ is an arbitrary generating function and $\ell^{\prime}(\tau)=\frac{\partial \ell}{\partial \tau}$.  Choosing $\ell(\tau)=\tau$ reproduces the free seed theories (Maxwell or Principal Chiral Model), while other choices yield Born–Infeld, ModMax, logarithmic, and $q$‐deformed interactions.

Moreover, one may construct root-$T\bar T$ and $T\bar T$ stress‐tensor operators in $d$ dimensions ~\cite{Babaei-Aghbolagh:2022uij,Babaei-Aghbolagh:2022leo, Ferko:2022cix,Conti:2018jho,Conti:2022egv,He:2025ppz},
\begin{eqnarray}
\mathcal{R}_\gamma
&=&\frac{1}{\sqrt{d}}\sqrt{T_{\mu\nu}T^{\mu\nu}-\tfrac{1}{d}T_\mu{}^\mu T_\nu{}^\nu},
\\
\mathcal{O}_\lambda
&=&\tfrac{1}{d}\bigl(T_{\mu\nu}T^{\mu\nu}-\tfrac{2}{d}T_\mu{}^\mu T_\nu{}^\nu\bigr),\nonumber
\end{eqnarray}
which, upon substitution of the general solution for $\hat{\mathcal{L}}(\alpha, \beta)$, collapse to the universal expressions
\begin{equation}\label{deformations}
\mathcal{R}_\gamma=\tau\ell^{\prime}(\tau),
\quad
\mathcal{O}_\lambda
=-\,\ell(\tau)\bigl(\ell(\tau)-2\tau\ell^{\prime}(\tau)\bigr).
\end{equation}
Thus, a single Courant–Hilbert framework provides a unified, dimension-independent classification of both duality-invariant electrodynamics and their two-dimensional integrable sigma-model counterparts.
\subsection*{New auxiliary field formalism }
In \cite{Russo:2025fuc}, a new auxiliary field formalism is introduced for self-dual nonlinear electrodynamics, governed by an auxiliary-field potential, which is highly different from the conventional Ivanov–Zupnik formalism~\cite{Ivanov:2002ab, Ivanov:2003uj} and more recent advances~\cite{Ferko:2024ali, Bielli:2024khq}. In this framework, causality guarantees a unique and consistent solution for the auxiliary field. A general solution for integrable models is encoded in the Lagrangian:
\begin{equation}\label{LAFF}
\hat{\mathcal{L}}(q_1,q_2)= - \frac{q_1}{y} + y \,q_2 -  \Omega(y),
\end{equation}
 where $y = e^\phi$ and $\phi$ is auxiliary field, and $\Omega(y)$ is a master potential determined by dualization constraints. The variation with respect to $y$ yields $\Omega^{\prime} (y)=\frac{q_1}{y^2} + q_2$, ensuring compatibility with the duality-invariant condition~\eqref{PDE4UV} and the integrability condition~\eqref{PDE222}.
 This formalism yields precise expressions for stress-energy tensors and their invariants, enabling the definition of marginal and irrelevant flow operators in terms of $\Omega(y)$. 
 Specifically, the root-$T\bar{T}$ and irrelevant $T\bar{T}$ operators are given by: \begin{equation}\label{lM2}
	\mathcal{R}_\gamma = y\, \Omega^{\prime} (y)\, , \quad \mathcal{O}_\lambda =- \Omega ^2(y) + y^2 {\Omega^{\prime}}^2 (y)\,.
\end{equation}
These expressions confirm that both types of deformations originate from a single universal potential and are seamlessly captured within the auxiliary-field formalism. This structure not only supports exact solvability and integrability but also ensures applicability across dimensions, unifying two-dimensional sigma models with four-dimensional duality-invariant electrodynamics under a shared deformation paradigm. 

\section*{Extended Coupling $(\lambda)$ to   pair space $(\lambda, \gamma)$ and generating $\gamma$-Flows}\label{gammm}
In this section, we present a systematic framework for constructing deformed theories in the (CH) approach and the new auxiliary field formulation. Our approach introduces novel field transformations that generalize the conventional single-coupling framework in two significant ways:
\begin{enumerate}
    \item Extension from a single coupling constant $\lambda$ to a two-parameter space $(\lambda, \gamma)$
    \item Preservation of the root-$T\bar{T}$ flow condition for all $\gamma$-coupled theories
\end{enumerate}
The fundamental transformation rule takes the general form:
\begin{equation}
    \mathcal{L}(\lambda) \rightarrow \tilde{\mathcal{L}}(\lambda, \gamma) = \mathcal{L}_{seed} + \lambda 
    \,\mathcal{L}_1 (\gamma)+ \lambda^2 \, \mathcal{L}_2(\gamma)+...,
\end{equation}
where $\mathcal{L}_{seed}$ is the Lagrangian for the seed theories in 2D or 4D and $\mathcal{L}_n(\gamma)$ is carefully constrained to maintain compatibility with the root-$T\bar{T}$ operator. The Lagrangian transformation $\mathcal{L}(\lambda) \rightarrow \tilde{\mathcal{L}}(\lambda, \gamma)$ is systematically implemented through coordinated transformations of the $\lambda$  dependent function, $\ell(\tau)$, potential, $\Omega(y)$, and the coupling $\lambda$.

\subsection*{Courant–Hilbert approach }
The most general integrable deformations follow from a Courant–Hilbert generating function admitting the weak‐field expansion \cite{BabaeiAghbolagh:2025rtt}
\begin{equation}
\ell^{\lambda}(\tau)=\tau+\lambda\,\mathcal{O}(\tau^2)+\lambda^2\,\mathcal{O}(\tau^3)+\cdots,
\end{equation}
which in the limit $\lambda\to0$ reproduces the 2D free PCM or 4D Maxwell seed. 
The marginal $\gamma$-flow corresponds to the universal flow equations in~\eqref{deformations} through rescalings of $\ell^{\lambda}(\tau)$ and the coupling $\lambda$, as:
\begin{equation}\label{TypI}
\ell^{(\lambda,\gamma)}(\tau)
= e^{\gamma}\,\ell^{\bar \lambda}(\tau),
\qquad
\bar \lambda=e^{\gamma}\,\lambda,
\end{equation}
defining a universal class of integrable theories where the deformation
\begin{equation}
\label{TypCH}
\partial_\gamma \hat{\mathcal{L}}^{(\lambda,\gamma)} = \tau\ell^{\prime}(\tau),
\end{equation}
preserves integrability via consistent transformations of the Lax structure while maintaining the root-$T\bar{T}$ operator structure.

\textbf{Uniform rescaling:}  A family arises from the simpler rescaling
\begin{equation}\label{TypII}
\ell^{(\lambda,\gamma)}(\tau)
= e^{\gamma}\,\ell^{\lambda}(\tau),
\end{equation}
so that in perturbation theory 
$\ell^{(\lambda,\gamma)}(\tau)=e^{\gamma}\tau+\lambda e^{\gamma}\mathcal{O}(\tau^2)+\cdots$. 
The corresponding Lagrangian of the CH-function \eqref{TypII} satisfies the $\gamma$-flow equation:
\begin{equation}\label{LLL}
\frac{\partial \hat{\mathcal{L}}^{(\lambda,\gamma)}}{\partial\gamma}
=\ell^{(\lambda,\gamma)}(\tau).
\end{equation}

\textbf{Single-trace flow:}  Finally, a single-trace deformation is implemented by 
\begin{equation}\label{TypIII}
\ell^{(\lambda,\gamma)}(\tau)
=\ell^{e^{\pm\gamma}\lambda}(\tau),
\end{equation}
leading to the  marginal flow:
\begin{equation}\label{SingleTR}
\frac{\partial\mathcal{L}^{(\lambda,\gamma)}}{\partial\gamma}
=\mp\frac1d\,T_\mu{}^\mu
=\mp\bigl(\ell^{(\lambda,\gamma)}(\tau)-\tau\ell^{\prime(\lambda,\gamma)}(\tau)\bigr). 
\end{equation}

These exhaust all marginal flows compatible with the integrability condition and encompass perturbative and exact closed‐form integrable theories. For the first time, we report a single-trace flow equation with respect to the $\gamma$ coupling in the context of Courant–Hilbert\footnote{If we consider the single-trace flow equation for the $\lambda$ coupling as $\frac{\partial\mathcal{L}^{(\lambda,\gamma)}}{\partial\lambda}
=-\frac{1}{d \, \lambda}\,T_\mu{}^\mu$, 
we can derive the single-trace flow equation~\eqref{SingleTR} by transforming~\eqref{TypIII}.}.
\subsection*{New Auxiliary field approach}
The transformed Lagrangian takes the form 
\begin{equation}\label{gamtrans}
 \Omega(y) \, \to \, \tilde{ \Omega}(\tilde{y})= e^{\gamma} \,\Omega( e^{- \gamma} y)\,, \qquad  \lambda \to e^{ \gamma}\, \lambda\,,
\end{equation}
where $\Omega(y)$ encodes the $\lambda$-dependence and $\tilde{ \Omega}(\tilde{y})$ is the modified potential with two parameters  $(\lambda, \gamma)$.
The potential transforms as $\Omega(y) \to \tilde{\Omega}(y,\gamma)$ with $\tilde{\Omega}(y,0) = \Omega(y)$ as the boundary condition. Based on the above conversions, it can be shown that the derivative of the potential $\tilde{ \Omega} (\tilde{y})$ satisfies the relation $\tilde{ \Omega}^{\prime} (\tilde{y})=\frac{q_1}{\tilde{y}^2} + q_2\,$. Applying the transformation rules established in \eqref{gamtrans}, we express the general modified Lagrangian with two coupling constants $(\lambda,\gamma)$ in the form:
\begin{equation}
  \tilde{  \mathcal{L}}(q_1,q_2, \tilde{y})= - \frac{q_1}{ \tilde{y}} +  \tilde{y} \,q_2 -  \tilde{ \Omega}(\tilde{y})\,.
\end{equation}
Transformations introduced in Eq.~\eqref{gamtrans} result in the formation of the root $T\bar{T}$-flow equation within the theory's Lagrangian, expressed as:
\begin{equation}\label{ARootTTbar}
\partial_\gamma \hat{\mathcal{L}}^{(\lambda,\gamma)} = \tilde{y} \, \tilde{ \Omega}^{\prime} (\tilde{y})\,.
\end{equation}

\section{Novel Integrable Theories}
In this section, we implement two previously established approaches to construct distinct classes of integrable theories characterized by the couplings $\gamma$ and $\lambda$. The (CH) approach generates both Born-Infeld and logarithmic theories~\cite{Russo:2024csd,Russo:2024dual,Russo:2024xnh}, while the new auxiliary field method produces $q$-deformed theories at $q = \frac{3}{4}$. This systematic framework yields novel families of closed-form integrable models, all of which remain consistent with the universal root $T\bar{T}$-flow equations introduced in the previous section.
\subsection{Theories from Courant–Hilbert approach}
\subsubsection{Generalized Born-Infeld theories }
We examine a canonical example of integrable models and invariant electromagnetic theories possessing Born-Infeld duality, characterized by the CH-function $\ell^{(\lambda)}(\tau)$ with coupling constant $\lambda$. This function is defined by the following expression:
\begin{equation}\label{lBI}
  \ell^{\lambda}(\tau) = -\frac{1}{\lambda}\left(1 - \sqrt{1 + 2\lambda \tau}\right).
\end{equation}
By applying the transformations in~\eqref{TypI} to the CH-function $\ell^{\lambda}(\tau)$ defined in~\eqref{lBI}, we obtain a modified CH-function incorporating both $\lambda$ and $\gamma$ coupling parameters. This generalized function generates the following Lagrangian:
\begin{equation}\label{Gqjqo}
    \hat{\mathcal{L}}_{GBI} = \frac{1}{\lambda}\left(-1 + \sqrt{1 - 2e^{-\gamma}\lambda q_1}\sqrt{1 + 2e^\gamma\lambda q_2}\right).
\end{equation}
Furthermore, the Lagrangian in Eq.~\eqref{Gqjqo} represents the first type-I theory that admits a closed-form expression~\cite{Bandos:2020hgy}. One can explicitly demonstrate that the root $T\bar{T}$ flow equation, given in Eq.~\eqref{TypCH}, holds for the Lagrangian defined in Eq.~\eqref{Gqjqo}.
We construct rescaled generalized Born-Infeld theories by applying transformations \eqref{TypII} to the CH-function in \eqref{lBI}. The rescaled Lagrangian reveals that the corresponding flow equation is equivalent to the CH function:
\begin{equation} \label{L0L}
\frac{\partial    \hat{\mathcal{L}}^{res}_{\text{GBI}}}{\partial \gamma} = \ell^{(\lambda,\gamma)}(\tau).
\end{equation}
Remarkably, this flow equation is energy-momentum independent, a novel result, as all previously known flow equations for such theories depended explicitly on the energy-momentum tensor. This represents a significant departure from conventional formulations in the literature.

The generalized Born-Infeld theories discussed here arise from applying the transformation
$\lambda^{\prime}= e^{\mp \gamma}  \lambda$,
(see~\eqref{TypIII}) to the CH-function~\eqref{lBI} , which we denote by $\hat{\mathcal{L}}^{ST}_{\text{GBI}}$. A key distinguishing feature of the resulting transformed theories is that, unlike the previous generalized Born-Infeld Lagrangians, their $\lambda \to 0$ limit reduces to either:
\begin{itemize}
    \item Maxwell's theory in four dimensions, or
    \item Principal Chiral Model (PCM) in two dimensions
\end{itemize}
The energy-momentum tensor derived for the Lagrangian $\hat{\mathcal{L}}^{ST}_{\text{GBI}}$, sector yields the trace:
\begin{eqnarray}\label{DLatgbi}
	{T_{\mu}}^{\mu}&=&  \frac{e^{\pm\gamma} \bigl(\sqrt{e^{\pm\gamma} - 2 \lambda q_1} -  \sqrt{e^{\pm\gamma} + 2 \lambda q_2}\bigr)^2}{ \lambda \sqrt{e^{\pm\gamma} - 2 \lambda q_1} \sqrt{e^{\pm\gamma} + 2 \lambda q_2}}\,.
\end{eqnarray}
One can explicitly verify that the $\gamma$-single trace flow equation in~\eqref{TypIII},  holds for the Lagrangian $\hat{\mathcal{L}}^{ST}_{\text{GBI}}$.
\subsubsection{Logarithmic theories}
We consider a class of two-dimensional integrable logarithmic field theories characterized by two independent coupling constants, $\gamma$ and $\lambda$. These models can be systematically derived through appropriate transformations of the fundamental equation~\eqref{TypI} when applied to the CH-function
\begin{equation}\label{CHlog}
    \ell(\tau) = -\frac{1}{\lambda}\log(1 -\lambda \tau)\,.
\end{equation}
The transformation procedure preserves the integrability structure while introducing the additional coupling parameter $\lambda$, which governs the logarithmic sector of the theory. This construction generalizes previous results by allowing for more flexible interaction terms in the Lagrangian while maintaining exact solvability.
The logarithmic  Lagrangians obtained from transformations \eqref{TypI}, are given by:
\begin{eqnarray}\label{LLog}
	\hat{\mathcal{L}}_{Log}&=&\frac{1}{\lambda}\bigg( -\sqrt{1 + 4 \lambda e^{-\gamma} q_1  - 4 \lambda^2 q_1 q_2} +1  \\
    &-&  \log\Bigl(\frac{e^{\gamma} }{2 \lambda q_1} \Bigl(1 -  \sqrt{1 + 4 \lambda e^{-\gamma} q_1 - 4 \lambda^2 q_1 q_2}\Bigr) \Bigr)\bigg)\nonumber.
\end{eqnarray}
It can be explicitly shown that the above Lagrangian satisfies the root $T\bar{T}$-flow equation
 \begin{equation}\label{rootLog1}
\frac{\partial \hat{\mathcal{L}}_{Log}}{\partial \gamma}=\tau \, {\ell^{\prime}}^{(\lambda , \gamma)}(\tau)\,.
\end{equation}
Applying rescaling~\eqref{TypII} to the CH-function~\eqref{CHlog}, we obtain a logarithmic Lagrangian $   \hat{\mathcal{L}}^{res}_{\text{Log}}$. This Lagrangian can be rescaled to recover the original form of~\eqref{LLog}. Furthermore, $   \hat{\mathcal{L}}^{res}_{\text{Log}}$ satisfies a flow equation with respect to the $\gamma$-coupling, denoted by $\frac{\partial  \hat{\mathcal{L}}^{res}_{Log} }{\partial \gamma}= \ell^{(\lambda , \gamma)} (\tau)$, which is proportional to the CH-function.  
By applying~\eqref{TypIII} transformations to the CH-function~\eqref{CHlog}, one can generate a logarithmic theory $\hat{\mathcal{L}}^{ST}_{\text{Log}}$ that satisfies the single-trace flow equation, analogous to the general Born theory of $\hat{\mathcal{L}}^{ST}_{\text{GBI}}$.
\subsubsection{q-Deformed Theories}
A large family of integrable $q$-deformed theories can be constructed for the CH-function $\ell(\tau)$, subject to the initial condition $\ell(\lambda = 0, \tau) = \tau$.
\begin{equation}\label{qDef}
    \ell(\tau) = \frac{1}{\lambda}\left(1 - \left(1 - \tfrac{1}{\mathbf{q}}\lambda\tau\right)^{\!\mathbf{q}}\right).
\end{equation}
These theories are deformed by the root-$T\bar{T}$ flow, as governed by the operator $\mathcal{R}_{\gamma}$ in the CH-function framework. For various values of the parameter \( q \), the PDE in \eqref{PDE222} can be solved by applying Eq.~\eqref{TypI} to Eq.~\eqref{qDef} within the (CH) approach. This includes special cases such as \( q = \tfrac{3}{2} \), \( q = \tfrac{3}{4} \), and the limiting configuration \( q \to \infty \).
\subsection{ Theories from new auxiliary field formalism}\label{3.3}
In this section, we study flow equations derived from the new auxiliary field approach, including a $q$-deformed theory and a novel theory.
\subsubsection{ $q$-deformed theories}\label{3.3}
We demonstrate that an appropriate potential $\Omega(y)$  new auxiliary field formalism reproduces the $q$-deformed theory in the CH approach for $q =3/4$ and yields the corresponding flow equations $\partial_\gamma \hat{\mathcal{L}}^{(\lambda,\gamma)} = \mathcal{R}_\gamma$ in~\eqref{lM2}. To achieve this, we consider the following potential:
\begin{equation}
\Omega(y)=\frac{1}{4\lambda}\Big(  (y^{-3} + 3 y)-4 \Big)\,,
\end{equation}
where $y$ is determined by:
\begin{equation}
y=\sqrt{\frac{2 \lambda q_1 +\sqrt{9 + 4 \lambda (\lambda q_1^2 - 3 q_2)}}{3 - 4 \lambda q_2}}\,.
\end{equation}
The Lagrangian takes the form \footnote{The Lagrangian~\eqref{q34} can be obtained exactly from the (CH) approach with the CH-function~\eqref{qDef} for $q=\frac{3}{4}$.}:
\begin{eqnarray}\label{q34}
  \mathcal{L}_{q=\frac{3}{4}}&=&\frac{1}{\lambda}- 2 q_1 \bigl(1 -  \frac{4\lambda }{9 }( \Delta - 2\lambda q_1^2+ 3 q_2)\bigr)^{1/4}\\
  & -& \frac{1}{\lambda} \bigl(1 -  \frac{4\lambda }{9 }( \Delta - 2\lambda q_1^2+ 3 q_2)\bigr)^{3/4}\nonumber\,,
\end{eqnarray}
with $\Delta$ defined as:
\begin{equation}
 \Delta=\sqrt{q_1^2 \bigl(9 + 4 \lambda^2 q_1^2 - 3 \lambda q_2)\bigr)}\,.
\end{equation}
Using the approach presented in~\eqref{gamtrans}, we can introduce the $\gamma$-coupling to the Lagrangian~\eqref{q34} such that the root flow equations take the form:
$
	\partial_\gamma \hat{\mathcal{L}}^{(\lambda,\gamma)} = \tilde{y}\, \tilde{\Omega}^{\prime} ( \tilde{y})\,  \, .
$
To achieve this, we apply \eqref{gamtrans} to the potential and redefine it as follows:
\begin{equation}
\tilde{ \Omega}(\tilde{y})=\frac{1}{4\lambda} (\frac{e^{3 \gamma}}{\tilde{y}^3} + \frac{3 \tilde{y}}{e^{\gamma}}-4 )\,,
\end{equation}
with the auxiliary field:
\begin{equation}
  \tilde{y}=\sqrt{\frac{2 \lambda q_1 + \sqrt{9 e^{2 \gamma} + 4 \lambda^2 q_1^2 - 12 e^{3 \gamma} \lambda q_2}}{3 e^{-\gamma} - 4 \lambda q_2}}\,.
\end{equation}
The generalized Lagrangian is:
\begin{equation}
  \hat{ \mathcal{L}}_{q=\frac{3}{4}}= \frac{1}{\lambda} + \sqrt{3 e^{-\gamma} - 4 \lambda q_2} \,\Bigl( \frac{e^{2 \gamma} ( 4 e^{\gamma} \lambda q_2-3 )}{\Sigma^3 \lambda}- \frac{2 q_1}{\Sigma}\Bigr)\,,
\end{equation}
where:
\begin{equation}
\Sigma =\sqrt{2 \lambda q_1 + \sqrt{9 e^{2 \gamma} + 4 \lambda^2 q_1^2 - 12 e^{3 \gamma} \lambda q_2}}\nonumber\,.
\end{equation}
The root $T\bar{T}$-flow equation in~\eqref{ARootTTbar} remains satisfied.

\subsection{ New Integrable Theory }\label{3.4}
We present here a novel integrable theory characterized by the potential $\Omega(y)$ and auxiliary variable $y$:
\begin{equation}
\Omega(y)= \frac{1}{\lambda} \Big(\frac{2 }{3 y} + \tfrac{1}{3} y^2-1\Big), \, \quad y=- \frac{\lambda}{2 }(\Lambda^{\frac{1}{3}} -  q_2 + \frac{q_2^2}{\Lambda^{\frac{1}{3}}}),
\end{equation} 
where $\Lambda$ is given by:
\begin{equation}
\Lambda= \frac{1}{\lambda^3} \Big(2 \sqrt{2 + 3 \lambda q_1) (2 + 3 \lambda q_1 + \lambda^3 q_2^3)}-4 - 6 \lambda q_1 -  \lambda^3 q_2^3 \Big)\nonumber.
\end{equation}
The Lagrangian for this new theory takes the form:
\begin{eqnarray}\label{newla}
 && \hat{\mathcal{L}}_{new}=\frac{1}{\lambda} \Big(1-  \tfrac{2  q_1}{q_2- \Lambda^{\frac{1}{3}}  -  \frac{q_2^2}{\Lambda^{\frac{1}{3}}}} +\frac{\lambda^2}{2 } q_2 (q_2- \Lambda^{\frac{1}{3}}  -  \frac{q_2^2}{\Lambda^{\frac{1}{3}}})\nonumber\\
  &-&\frac{2 }{3\lambda ( q_2- \Lambda^{\frac{1}{3}}  -  \frac{q_2^2}{\Lambda^{\frac{1}{3}}})} \bigl(2 + \frac{1}{8 }\lambda^3( q_2- \Lambda^{\frac{1}{3}}  -  \frac{q_2^2}{\Lambda^{\frac{1}{3}}})^3\bigr)\Big).
\end{eqnarray}
By applying the transformation~\eqref{gamtrans}, we introduce the $\gamma$-coupling into the new Lagrangian~\eqref{newla}, allowing us to express the root $T\bar{T}$ flow as a (PDE) in terms of the auxiliary field 
$\tilde{y}$.
The modified potential  is 
\begin{eqnarray}
\tilde{\Omega}(\tilde{y})=\frac{1}{\lambda} \Big(\frac{2 e^{\gamma}}{3  y} + \tfrac{e^{-2\gamma}}{3} y^2-1\Big)\,,
\end{eqnarray}
with the auxiliary variable $\tilde{y}$ expressed through:
\begin{equation}
\tilde{y}=- \tfrac{1}{2} e^{\gamma} \tilde{ \Lambda}^{1/3} + \tfrac{1}{2} e^{2 \gamma} \lambda q_2 -  \frac{e^{3 \gamma} \lambda^2 q_2^2}{2 \tilde{ \Lambda}^{1/3}}\,,
\end{equation} 
where $\tilde{ \Lambda}$ is given by:
\begin{eqnarray}
\tilde{ \Lambda}&=&-4 - 6 e^{-\gamma} \lambda q_1 -  e^{3 \gamma} \lambda^3 q_2^3 \\
&+& 2 \sqrt{(2 + 3 e^{-\gamma} \lambda q_1) (2 + 3 e^{-\gamma} \lambda q_1 + e^{3 \gamma} \lambda^3 q_2^3)}\nonumber\,.
\end{eqnarray}
This leads to an explicit expression for the generalized Lagrangian 
\begin{eqnarray}\label{newlagamma}
\tilde{\mathcal{L}}_{new}&=&\frac{1}{\lambda} + \frac{3 e^{-\gamma} q_1}{\tilde{ \Lambda}^{1/3}} -  \tfrac{1}{4} e^{\gamma} \tilde{ \Lambda}^{1/3} q_2 + \tfrac{1}{4} e^{2 \gamma} \lambda q_2^2 \\
  &- & \frac{e^{3 \gamma} \lambda^2 q_2^3}{4 \tilde{ \Lambda}^{1/3}} +\frac{2 \tilde{ \Lambda}^{1/3} + 3 \lambda^2 q_1 q_2 -  \frac{3 e^{\gamma} \lambda^3 q_1 q_2^2}{\tilde{ \Lambda}^{1/3}}}{\tilde{ \Lambda}^{2/3} \lambda -  e^{\gamma} \tilde{ \Lambda}^{1/3} \lambda^2 q_2 + e^{2 \gamma} \lambda^3 q_2^2}\nonumber\,,
\end{eqnarray}
fully capturing the dynamics of the $\lambda$ and $\gamma$ deformations.
An expansion in $\lambda$ reveals the perturbative structure:
\begin{eqnarray}
  \tilde{ \mathcal{L}}_{new}&=&- e^{-\gamma} q_1 + e^{\gamma} q_2 + \frac{e^{-2 \gamma}}{4}  \lambda (q_1 + e^{2 \gamma} q_2)^2\\
  & +&  \frac{e^{-3 \gamma} }{12} \lambda^2 (-2 q_1 + e^{2 \gamma} q_2) (q_1 + e^{2 \gamma} q_2)^2\nonumber\\
  & +& \frac{ e^{-4 \gamma}}{48}\lambda^3  (q_1 + e^{2 \gamma} q_2)^2 (7 q_1^2 - 4 e^{2 \gamma} q_1 q_2 + e^{4 \gamma} q_2^2)\,. \nonumber
\end{eqnarray}
This new Lagrangian in~\eqref{newlagamma} satisfies the root $T\bar{T}$-flow in~\eqref{ARootTTbar}.

\section*{Conclusions}
This work establishes a systematic correspondence between duality-invariant nonlinear electrodynamics (NED) in four dimensions and integrable sigma models in two dimensions. The connection is realized through the Courant–Hilbert framework and a novel auxiliary field formulation. A central insight is that both systems are governed by a common partial differential equation, arising from the self-duality constraint in 4D NED and the integrability condition in 2D sigma models, respectively. This structural parallel reveals a profound geometric and algebraic link between self-dual electricmagentic theories and 2D integrable models.

By extending traditional deformation techniques to include both  ($\gamma$) and ($\lambda$) couplings, we show that the root-$T\bar{T}$ and $T\bar{T}$ deformations naturally emerge as complementary flows within a unified geometric setup. The developed framework leads to several key outcomes:

\begin{itemize}
    \item A unified method for constructing integrable two-dimensional field theories and $T\bar{T}$ flows,
    \item Generalizations of known models (e.g., Born-Infeld and $q$-deformed) through an additional $\gamma$ coupling,
    \item A novel integrable theory with an explicit closed-form Lagrangian,
\end{itemize}

Explicit closed-form Lagrangians are derived for a range of integrable theories, including generalized Born-Infeld, logarithmic, and $q$-deformed models.
A general perturbative Lagrangian, derived using the (CH) approach, can be found in the appendix. This perturbative Lagrangian generates all theories up to $(\lambda^3)$.
All are unified via a single generating function $\ell(\tau)$, capturing marginal deformations in a compact form and significantly enriching the space of exactly solvable models.

The auxiliary field method provides a powerful and systematic approach to analyzing integrable deformations. Notably, it reveals how the inclusion of $\gamma$ leads to consistent integrability-preserving flows, characterized by Lax representations and controlled deviations from classical deformation patterns.

A central finding is the universality of the root-$T\bar{T}$ flow, encapsulated by the equation:
\begin{equation}
\partial_\gamma \mathcal{L} = \mathcal{R}_\gamma\,,
\end{equation}
which holds across all models under the new auxiliary field formalism. This suggests an underlying geometric principle that governs the behavior of these flows, independent of specific Lagrangian realizations.

The new integrable model introduced here exhibits a distinctive interaction structure, with a perturbative hierarchy that differs from both Born-Infeld and $q$-deformed theories. Its exact form invites further quantum investigation on the quantum aspects.

\begin{acknowledgments}
We are grateful to Jue Hou, J. G. Russo, and Dmitri Sorokin for their interest in this work and the fruitful discussions that followed. The work of H.B.-A. was conducted as part of the PostDoc Program on {\it Exploring TT-bar Deformations: Quantum Field Theory and Applications}, sponsored by Ningbo University. This research was partly supported by NSFC Grant No.11735001, 12275004, 12475053, and 12235016.
\end{acknowledgments}

\bibliography{referen}
%
\renewcommand{\thesubsection}{{S.\arabic{subsection}}}
\setcounter{section}{0}
\section*{Supplemental material}

\subsection{Generalized perturbative theory}\label{222.11} 
For general theories, the CH function can incorporate higher-order terms in $\tau$ with coupling to $\lambda$. A notable property of the function $\ell(\tau)$ is its power series expansion around $\lambda = 0$. We consider a general perturbative expansion for the CH function as follows:
\begin{equation}\label{Lexpand}
	\ell^{(\lambda , \gamma)} (\tau) =\sum_{i=0}^\infty \,n_i\,  \lambda^i\, f^{\mathfrak{I}}_i(\gamma)\, \tau^{i+1} 
\end{equation}
where the $n_i$ unfixed constants are theory-dependent and $\mathfrak{I} = I, II, III$ corresponds to the three type of transformations: Type-I~\eqref{TypI}, Type-II~\eqref{TypII}, and Type-III~\eqref{TypIII}.
We can show that for the triple-type $\gamma$-flow equations to hold, the $\gamma$-functions $f^{\mathfrak{I}}_i(\gamma)$, must satisfy the following relations:  
\begin{equation}\label{FI}
f^{I}_i(\gamma)= e^{(i+1) \gamma} ,   \quad f^{II}_i(\gamma)=   e^{ \gamma} , \quad f^{III}_i(\gamma)=  e^{ \pm i \gamma}, 
\end{equation}
By considering the CH-function~\eqref{Lexpand} and substituting the $\gamma$-functions $ f^{\mathfrak{I}}_i(\gamma)$ from Eq.~\eqref{FI}, we obtain the generalized perturbative Lagrangian in three types. For the case $n_0=1$, these three Lagrangian of the order of $\lambda^3$ in terms of the two variables $q_1$ and $q_2$ as follows:
\begin{itemize}
    \item Generalized type-I Lagrangian corresponding to the root $T\bar{T}$-flow equation:
\begin{eqnarray}\label{G1}
&&\hat{\mathcal{L}}^I (q_1,q_2)= e^{\gamma} q_2 - e^{-\gamma} q_1+ e^{-2 \gamma} n_1 \lambda (q_1 + e^{2 \gamma} q_2)^2 \nonumber\\
    &+& \lambda^2 \bigl( e^{-3 \gamma} n_2 (q_1 + e^{2 \gamma} q_2)^3-4 e^{-3 \gamma} n_1^2 q_1 (q_1 + e^{2 \gamma} q_2)^2 \bigr)\nonumber \\
    &+& \lambda^3 \bigl( e^{-4 \gamma} n_3 (q_1 + e^{2 \gamma} q_2)^4 -12 e^{-4 \gamma} n_1 n_2 q_1 (q_1 + e^{2 \gamma} q_2)^3  \nonumber\\
    &+& 8 e^{-4 \gamma} n_1^3 q_1 (q_1 + e^{2 \gamma} q_2)^2 (3 q_1 + e^{2 \gamma} q_2)\bigr)
\end{eqnarray}
 \item The generalized Type~II and Type~III Lagrangians for the uniform rescaling flow and the $\gamma$-single trace flow equations are obtained by applying the functions $f^{II}_i(\gamma)$ and $f^{III}_i(\gamma)$, respectively.
\end{itemize}
All novel theories studied in this work, to arbitrary order in $\lambda$, can be derived from the CH function expansion in~\eqref{Lexpand}. By selecting appropriate coefficients $\{n_i\}$, we can construct the CH function expansion up to $\mathcal{O}(\lambda^3)$. 
As shown in Table~\eqref{T11}, these novel theories form a special subgroup of the general Lagrangian~\eqref{G1}, when suitable coefficients $\{n_i\}$ are chosen.
\begin{center}
\begin{tabular}{|c|c|c|c|}
\hline
		Models & $\lambda^1$-Order & $\lambda^2$-Order & $\lambda^3$-Order    \\ \hline
		GBI &    $n_1=-\frac{1}{2}$& $n_2=\frac{1}{2}$& $n_3=-\frac{5}{8}$  \\  \hline
		Logarithmic&   $n_1=\frac{1}{2}$& $n_2=\frac{1}{3}$& $n_3=\frac{1}{4}$  \\  \hline
		$q=3/2$ &  $n_1=-\frac{1}{6}$& $n_2=-\frac{1}{54}$& $n_3=-\frac{1}{216}$   \\ \hline
		$q=3/4$ &  $n_1=\frac{1}{6}$& $n_2=\frac{5}{54}$& $n_3=\frac{5}{72}$   \\ \hline
		$q \to \infty$ &   $n_1=-\frac{1}{2}$& $n_2=\frac{1}{6}$& $n_3=-\frac{1}{24}$   \\ \hline
        
		New Theory&   $n_1=\frac{1}{4}$& $n_2=\frac{1}{12}$& $n_3=\frac{1}{48}$   \\ \hline
\end{tabular}
\captionof{table}{ \label{T11}  Novel theories  via General perturbative Lagrangians}
\end{center}

\end{document}